\documentstyle[12pt,aaspp4]{article}
\def\sun {$_{\scriptscriptstyle \odot}$}

\begin{document}

\title{Helium Star/Black Hole Mergers: A New Gamma-Ray Burst Model}
\author{C. L. Fryer \& S. E. Woosley}

\affil{Department of Astronomy and Astrophysics, University of
California, Santa Cruz, CA 95064}

\authoraddr{Board of Studies in Astronomy and Astrophysics, University
of California, Santa Cruz, CA 9506}

\authoremail{cfryer@ucolick.org, woosley@ucolick.org}

\begin {abstract}
We present a model for gamma-ray bursts (GRB's) in which a stellar
mass black hole acquires a massive accretion disk by merging with the
helium core of its red giant companion.  The black hole enters the
helium core after it, or its neutron star progenitor, first
experiences a common envelope phase that
carries it inwards through the hydrogen envelope.  Accretion of the
last several solar masses of helium occurs on a time scale of roughly
a minute and provides a neutrino luminosity of approximately 10$^{51}$
- 10$^{52}$ erg s$^{-1}$. Neutrino annihilation, 0.01\% to 0.1\%
efficient, along the rotational axis then gives a baryon loaded
fireball of electron-positron pairs and radiation (about 10$^{50}$ erg
total) whose beaming and relativistic interaction with circumstellar
material makes the GRB (e.g., Rees \& Meszaros 1992). The
useful energy can be greatly increased if energy can be extracted from
the rotational energy of the black hole by magnetic interaction with
the disk.  Such events should occur at a rate comparable to that of
merging neutron stars and black hole neutron star pairs and may be
responsible for long complex GRB's, but not short hard ones.
\end {abstract}

\keywords{gamma rays: bursts, gamma rays: theory, black hole physics} 
\section{INTRODUCTION}

Many close, massive binaries are expected to pass through an evoutionary
phase during which a compact object (neutron star or black hole) enters the
envelope of its red giant companion. Some of these systems eject the hydrogen
envelope and eventually evolve into doubly degenerate binaries such as the
Hulse-Taylor pulsar system (Hulse \& Taylor 1975, van den Heuvel
1995).  In others, however, the orbital
energy of the system is insufficient to remove the hydrogen envelope
and thus shut off drag before the compact object enters the helium core.  As
the compact object and helium core coalesce, the helium core is tidally
disrupted into an accretion disk around the compact object with a radius
equal to a fraction of the initial helium core, $\sim$10$^{9} - 10^{10}$ cm.
The accretion rate onto the compact object may be initially as large as the
Bondi-Hoyle rate, almost a solar mass per second (Houck \& Chevalier 1991;
Brown 1995; Chevalier 1996; Fryer, Benz, \& Herant 1996).  Later, following
disk formation, the accretion rate is limited by the viscous time scale,
$(\alpha \Omega)^{-1} \sim 100$ s, where $\alpha$ is the disk viscosity
parameter, typically 0.1, and $\Omega$ is the orbital angular velocity.
During the merger process, the compact object becomes a black hole, if it was
not already.

The resultant object, a $\sim 3$ M\sun black hole accreting the
remaining several solar masses of helium, resembles, except for the larger
store of available mass and higher specific angular momentum, $j \sim
10^{18}$ cm$^2$ s$^{-1}$, the class of accretion-disk/black-hole GRB models
based on the merger of neutron star pairs (NS/NS) or black hole neutron star
(BH/NS) pairs. It also resembles, except for its larger angular momentum,
rotating failed supernovae (Woosley 1993). Because of the large angular
momentum, the accreting black hole will likely become a Kerr hole with a
rotational speed near the maximum allowed by relativity.  As the black hole
spins up, the efficiency at which the gravitational energy is released rises
from 0.06 for Schwarzschild (non-rotating) geometries to as much as 0.42
$\dot M c^2$ for Kerr geometries.  Taking 0.1 as a
representative value of this efficiency, the accreting black hole's neutrino
luminosity for 100 seconds is about $10^{51}-10^{52}$ erg s$^{-1}$.
Conversion of 0.01 to 0.1\% of this energy (see below) into a pair fireball
by neutrino annihilation along the rotational axis leads to burst energies in
the range of 10$^{49}$ - 10$^{51}$ erg, equal to those achieved by NS/NS
mergers (Ruffert et al. 1997).  The strong beaming that will likely accompany
accretion-disk/black-hole GRB models then allows sufficiently-high effective
energies to power the GRB.

As we shall also see, the rate of such helium mergers may be an 
order of magnitude greater than that of NS/NS and BH/NS binaries combined.
GRB's from helium mergers differ from those of NS/NS and BH/NS mergers
in that no appreciable gravitational wave signal should precede a
helium merger gamma-ray burst as they would for gamma-ray bursts from
the merger of NS/NS or BH/NS binaries.  In addition, gamma-ray bursts
from helium mergers occur immediately at their star-formation sites,
whereas the delay in the merger of NS/NS and BH/NS binaries allows
these gamma-ray burst progenitors to leave not only the site of their
formation before merging, but, in some cases, their host galaxy as
well.

\section {PRE-BURST EVOLUTION}

Our model begins with a close binary system of two massive stars (each
greater than 8 M\sun).  As the more massive star (primary) evolves off the
main sequence, it fills its Roche-lobe and transfers mass onto its companion
(secondary).  If the system remains bound after the supernova explosion of
the primary, a binary composed of a neutron star and a massive main-sequence
star results, possibly a massive X-ray binary.  When the secondary, in turn,
evolves off the main sequence and overflows its Roche lobe, the neutron star
enters a common envelope with the secondary and begins to spiral towards the
secondary's helium core.  The inward motion continues for roughly 1
(Sandquist et al. 1998) to 1000 years 
(Taam, Bodenheimer, \& Ostriker 1978) until either the
secondary's hydrogen envelope is completely ejected and the system becomes a
helium star/neutron star binary or the compact star merges with the helium
core.

Whether or not the neutron star then finally merges with its secondary
companion depends upon one of the most uncertain parameters in population
synthesis studies: the common envelope efficiency ($\alpha_{\rm CE}$).  This
single parameter describes the efficiency at which the orbital energy,
when injected into the companion's hydrogen atmosphere as the 
neutron star inspirals during a common envelope phase, drives off 
the companion's envelope.  
Using $\alpha_{\rm CE}$, one can estimate the final orbital separation
($A_f$) in terms of the initial separation ($A_i$) of the binary after common
envelope evolution (Webbink 1984): \begin{equation}
\frac{A_{f}}{A_{i}}=\frac{\alpha_{\rm CE}\, r_{L}\, q}{2} \left(\frac{M_{He}}
{M_{s}-M_{He} + \frac{1}{2}\alpha_{\rm CE} r_{L} M_{NS}} \right),
\end{equation} where $M_{s}$, $M_{He}$, and $M_{NS}$ are, respectively, the
masses of the secondary, the secondary's helium core, and of the neutron star
and $r_{L}=R_L/A_i$ is the dimensionless Roche lobe radius of the secondary.
Some hydrodynamical models have simulated this inspiral, but only for
specific systems.  Although a general consensus of the value of the common
envelope efficiency has not been achieved, best estimates give $\alpha_{\rm
CE} \sim 0.5$ (Taam et al. 1997). 

During the inspiral, the neutron star accretes material at the
Bondi-Hoyle rate, releasing the accretion energy via neutrino emission
(Houck \& Chevalier 1991, Chevalier 1993, Chevalier 1996, 
Brown 1996; Fryer, Benz, \& Herant 1996).  Bethe \& Brown (1998) 
have calculated that in the hydrogen inspiral alone, a neutron star 
is likeley to accrete $\sim 1 M_{\odot}$, leading to the collapse 
of that neutron star into a black hole.  In the tenuous layers of 
hydrogen envelope, Chevalier (1996) showed that if the infalling 
material had sufficient angular momentum, the temperature of the 
material on the neutron star surface may not get high enough 
to emit neutrinos, and the Eddington limit would constrain 
the accretion rate.  However, near or below the surface of the 
helium star, where most of the accretion occurs, the densities 
(and the subsequent Bondi-Hoyle accretion rates) are so high 
$> 1M_\odot$\,/yr, that angular momentum will not prevent 
super-Eddington accretion via neutrino emission (Chevalier 1996).  
The analysis of Chevalier assume that all of the mass accretes
onto the neutron star flows through the disk, but accretion 
along the angular momentum axis dominates the flow (Fryer \& 
Kalogera 1998).  In addition, 3-dimensional simulations of Bondi-Hoyle 
accretion suggest that the amount of angular momentum accreted 
accreted in common envelopes may be much less than that 
assumed by Chevalier (e.g. Ruffert \& Anzer 1995, Ruffert 1997), 
again weakening the effects of angular momentum.
Thus, for all of these reasons, the Bethe \& Brown (1998) 
estimate of the accreted mass is valid.

As the compact object merges with the helium core, the Bondi-Hoyle 
accretion rate in the helium core can reach $\sim 1M_\odot/$s.  
From population synthesis calculations using the code developed 
by Fryer, Burrows, \& Benz (1998), the helium core and compact 
object have average masses of $\sim 4 M_\odot, 2 M_\odot$ 
respectively at the time of merger, and the mean
specific angular momentum of the system is $j \approx M_{\rm BH}
M_{\rm He} \sqrt{G R_{\rm He}/(M_{\rm BH}+M_{\rm He})^3} \sim 10^{18}
{\rm cm}^2/{\rm s}$ where $G$ is the gravitational constant and
$M_{\rm BH}$ is the black hole mass when it reaches the helium core
radius ($R_{\rm He}$).  As the two objects coalesce, the orbital
energy will drive off what remains of the hydrogen envelope as well as
some of the helium core.  The angular momentum will be injected into
the system to form a rotating disk of helium around the black hole.

\section {MAKING THE BURST}

Whether or not the resultant system is a feasible model for gamma-ray
bursts depends upon the accretion rate of the disk onto the black
hole.  Based upon the $\nu \bar \nu$ annihilation paradigm for black
hole-disk GRB models (Goodman, Dar, \& Nussinov 1987; Paczynski 1991;;
Woosley 1993; Ruffert et al. 1997), we can estimate the energy
produced in electron-positron pairs for our model.  For thin disk accretion,
the efficiency at which gravitational potential energy is emitted in
radiation (neutrinos) is well known (e.g, Shapiro \& Teukolsky 1983):
5.7\% for Schwarzchild black holes, and 42.3\% for maximally rotating 
Kerr black holes.  The black hole not only gains mass from the
accreting material, but also angular momentum, roughly equal to the 
angular momentum of the material at the last stable orbit.   
The black hole will reach its maximal rotation after accreting 
$\Delta M = 1.846 M_{\rm BH}$ for thin-disk accretion (Thorne 1974),
or $\Delta M \approx M_{\rm BH}$ for thick-disk accretion 
(Abramowicz \& Lasota 1980).  In the helium merger model, 
$\sim 4-5 M_\odot$ accrete onto the $1.4 M_\odot$ progenitor 
neutron star/black hole.  The early accretion will accrete 
along the angular momentum axis, but enough may accrete along 
the equator to spin-up the black hole, and gravitational 
energy conversion efficiencies may be as high as 10-20\% 
for these models.

This neutrino emission, if emitted in a disk, is then converted into
electron-positron pairs with an efficiency (Ruffert et al. 1997):
\begin{equation}
L_{\rm pair} \approx 3 \times 10^{46}  \left(\frac{L_{\nu}}{10^{51} {\rm
erg/s}}\right)^2 \left( \frac{<\epsilon_{\nu}>}{13 {\rm MeV}}\right)
\left(\frac{20 {\rm km}}{R_d} \right) \, {\rm erg \, s^{-1}}
\end{equation} 
where $L_{\nu}$ is the total neutrino luminosity (all flavors), 
$<\epsilon_{\nu}>$ is the mean neutrino energy, and $R_d$ is 
the inner disk radius.  To estimate a maximum energy in 
electron-positron pairs which will then drive the  GRB, we 
assume an accretion rate equal to the Bondi-Hoyle 
rate ($1 M_\odot/$s) for a Kerr black-hole 
accreting $1 M_\odot$:  $\sim 10^{52}$erg.   However, for a
Schwarzchild black hole accreting  $1 M_\odot$ at accretion disk rates of
$M_{\rm disk}/(\alpha \Omega)  = 0.01 M_\odot/$s, the energy drops to
$10^{48}$erg.  Typical  values for a $4 M_\odot$ accretion disk scenario lead
to  energies between $10^{49}-10^{50}$erg, quite comparable to  the burst
energies of NS/NS mergers (Ruffert et al. 1997).

A more promising model relies upon a strong magnetic field 
being produced in the accretion disk, which can then tap the 
rotational energy of the black hole to power a GRB 
(Blandford \& Znajek 1977; MacDonald et al. 1986; Paczynski 1991,
1997; Woosley 1993; Rees \& Meszaros 1997, Katz 1997):
\begin{equation}
L_{\rm rot}=10^{50} \left ( \frac{j c}{G M_{\rm BH}}\right )^2 
\left ( \frac{M_{\rm BH}}{3 M_\odot}\right )^2 
\left ( \frac{B}{10^{15} G}\right )^2 \, {\rm erg\,s^{-1}}
\end{equation}
where $j$ is the specific angular momentum of the black hole and $B$
is the magnetic field strength in the disk.  The large disks of our
helium merger model will spin-up the black hole, providing sufficient
energy and power for strongly magnetic fields to produce more
than $10^{52}$ erg of beamed energy over 100 s provided that a fraction of
order 1\% of the equipartition field value, $B^2/8 \pi \sim \rho v^2$ is
attained in the inner disk. 

\section {EVENT RATES}

Our burst energy estimates are well within limits to drive a GRB, especially
if the strong beaming, which is likely to occur, is included.  But for helium
mergers to be a viable gamma-ray burst model, they must also have a
sufficiently high formation rate to explain the observations.  
Population synthesis studies of massive binaries is fraught with 
a variety of unknown parameters:  e.g., kicks imparted onto neutron 
stars at birth, common envelope efficiency ($\alpha_{\rm CE}$), 
the initial mass function, the binary mass ratio distribution.  
Even the stellar radii during giant phases are not known to 
accuracies better than a factor of $\sim 2-5$.  Using a slightly 
modified version of the Monte Carlo code described in 
Fryer, Burrows, \& Benz (1998), we have run a series of 
population synthesis calculations (Fryer, Woosley, \& Hartmann 1998).    
Here we present specific results of two simulations using 
two delta-function kick magnitudes (50,150\,km\,s$^{-1}$) directed 
isotropically\footnote{We set $\alpha_{\rm CE}=0.5$, we use a Scalo 
(1986) initial mass function and a flat mass ratio distribution.  
See Fryer, Burrows, \& Benz (1998) and Fryer, Woosley, \& Hartmann (1998) 
for details.}.

In Figure 1, we compare population synthesis results of the formation 
rate vs. age of gamma-ray bursts from helium-mergers with that from 
NS/NS and BH/NS binaries combined for a galaxy with a burst of star 
formation and a galaxy with a constant supernova rate of 
$10^{-2}$\,y$^{-1}$.  These rates depend sensitively on the supernova 
kick and many of the binary parameters which may alter the formation
rate by over an order of magnitude (Fryer, Burrows, \& Benz 1998; 
Fryer, Woosley, \& Hartmann 1998), but the rate for helium mergers remains 
comparable to, and generally greater (often by an order of magnitude) 
than, the gamma-ray burst rate from NS/NS and BH/NS binaries
combined.  The rate is enough to provide the observed bursts 
even if a large beaming factor is invoked (Wijers et al. 1998).

Whereas helium merger GRB's tend to occur in the star
formation regions in which they are born, NS/NS and BH/NS binaries 
may not merge until long after their creation and may, therefore, 
leave these regions before exploding as a GRB.  Many NS/NS and BH/NS 
binaries form with systemic velocities which not only drive them out 
of their places of birth, but may also drive them beyond their host 
galaxies.  Assuming that the host galaxy's gravitational potential 
has no effect on the outmoving binary (systemic velocities are 
$\sim 100-200$\,km\,s$^{-1}$, so this assumption is valid for 
low-mass galaxies), we can estimate the distribution of 
distances these binaries travel prior to merging as a GRB (Figure 2).

The dominant factor governing the distance estimates is the 
merger timescale for the binaries.  This timescale, in turn, 
depends sensitively upon the relation of the helium star 
radius with its mass.  For our simulations we use 
following helium star mass-radius relation\footnote{This 
relation uses a radius scaled down by a factor of 5 from the 
mass-radius relation derived in Kalogera \& Webbink (1998) 
to match the radii from Woosley, Langer, \& Weaver (1995)}:
\begin{equation} 
\mbox{log} R_{\rm He,max}=
\left\{ \begin{array}{ll} 2.398-2.013 \, \mbox{log} \, M_{\rm He} &
M_{\rm He} \leq 2.5
\mbox{M}_{\odot} \\ -0.699+0.0557 \,(\mbox{log} \, M_{\rm He} - 0.172)^{-2.5}
& M_{\rm He} >
2.5
\mbox{M}_{\odot}
.  \end{array} \right. 
\end{equation} 
For those compact objects whose inspiral take them within the helium 
star radius, the system merges and becomes a helium merger 
GRB, not a NS/NS or BH/NS binary.  Hence, 
to form a NS/NS or BH/NS binary, before the supernova explosion 
of the helium star, the helium star/compact object separation 
must be greater than this radius.  This limiting pre-explosion 
separation defines the orbital separation distribution after 
the supernova explosion of the helium star.  As the helium 
star radius decreases, so then does the mean orbital separations 
of the NS/NS and BH/NS binaries, and hence, the merger times 
and distances traveled before merger decrease as well.  
In figure 2, we see that the mean distance traveled at the 
time of merger is roughly $\sim10-100$\,kpc.  If the helium star 
radii were a factor of 5 higher, this mean would increase 
to $\sim 1$\,Mpc.  Thus, it is possible that GRB's driven 
by the merger of compact binaries may not occur in their 
host galaxy, whereas helium mergers will all occur therein.

\acknowledgements 

The authors appreciate stimulating conversations with Hans Bethe and
Gerry Brown regarding common envelope evolution and black hole
formation.   We'd also like to thank Peter Bodenheimer for his advice 
on neutron star inspiral calculations, Bob Popham for sharing with us
output from his accretion disk calculations, and Alex Heger for 
helpful suggestions.  This work was supported by NASA (NAG5 2843 and MIT SC
A292701) and the NSF (AST 94-17161).

\clearpage

\begin{figure}
\plotfiddle{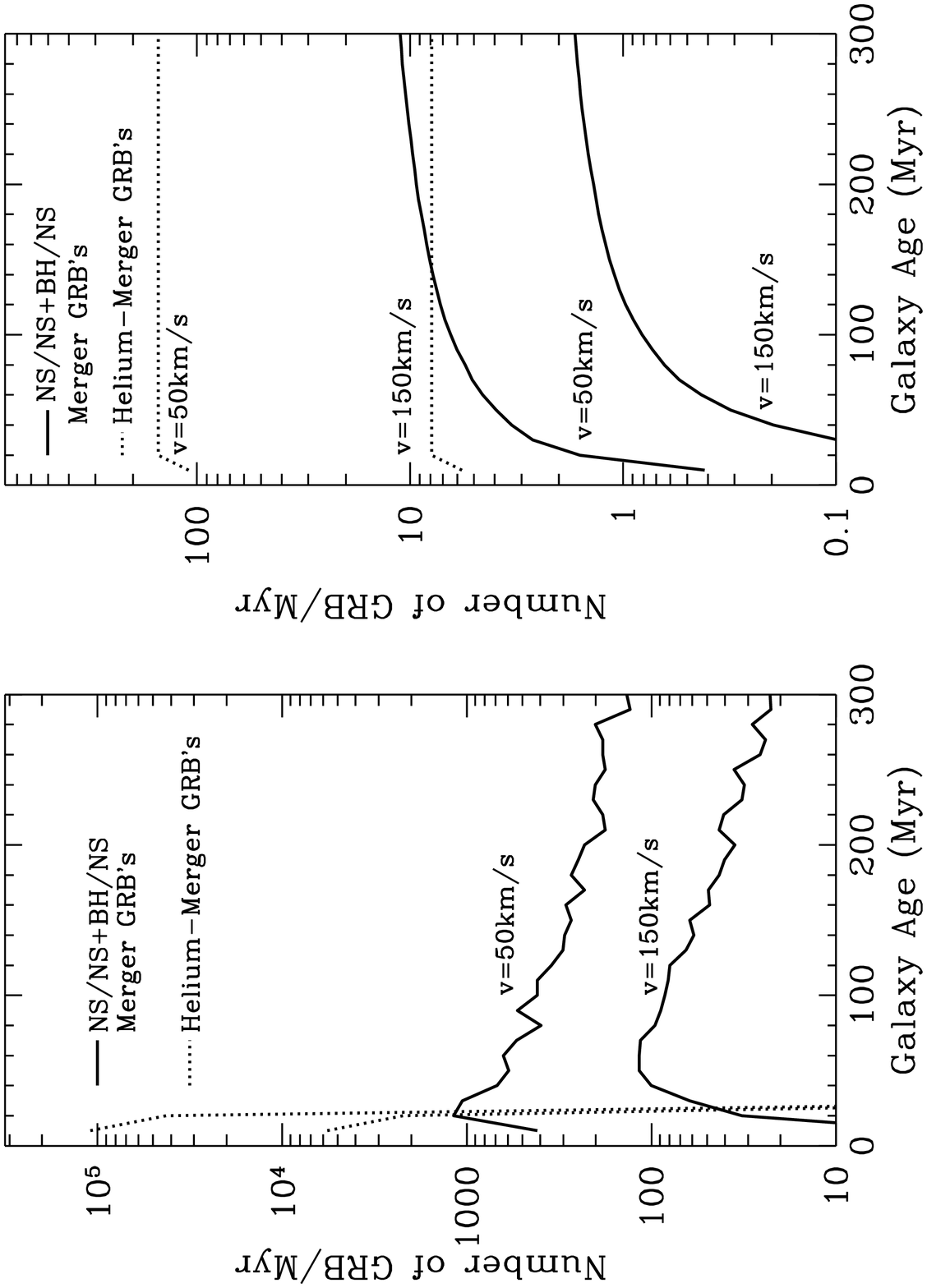}{7in}{-90}{70}{70}{-280}{520}
\caption{(a) The number of gamma-ray
bursts vs. age for a single burst of star formation with
$10^8$ supernovae.  Gamma-ray bursts continue to occur from mergers of
NS/NS and BH/NS binaries long after the initial burst of star
formation.  Helium merger gamma-ray bursts occur along with the star
formation and should not be observed in old stellar populations.  (b)
The number of gamma-ray bursts produced per Myr vs. age
assuming a constant supernova rate of 0.01 y$^{-1}$.  The helium
merger gamma-ray burst rate quickly reaches a peak and then remains
flat.}
\end{figure}

\begin{figure} 
\plotfiddle{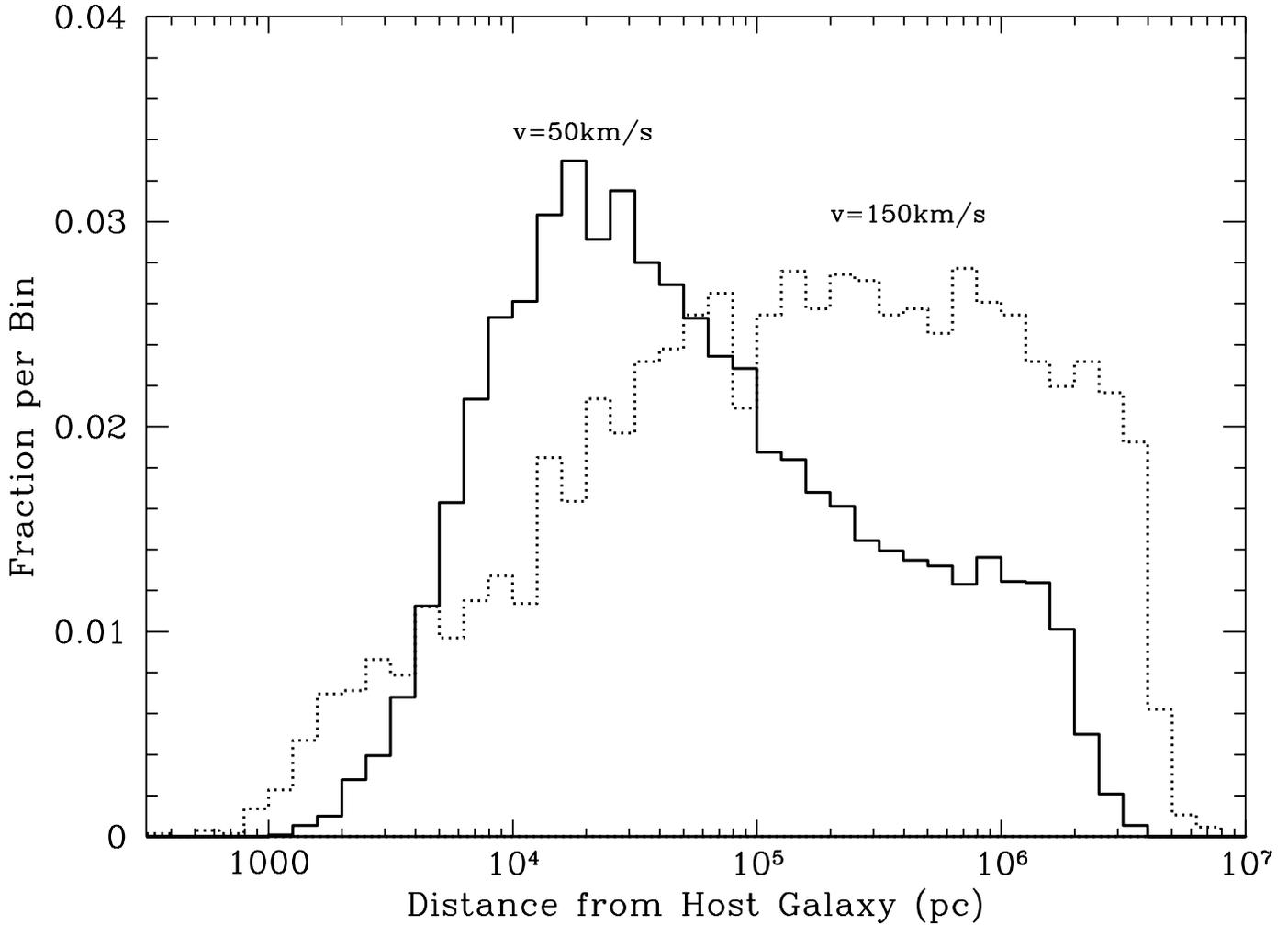}{7in}{-90}{70}{70}{-280}{520}
\caption{Gamma-Ray Burst Location.  Distribution of gamma-ray
bursts vs.  distance from formation site assuming no effects from the
galactic potential for the NS/NS and BH/NS binary mechanism.  Helium
merger gamma-ray bursts will explode in the star formation region in
which they were produced whereas some gamma-ray bursts from degenerate
binaries may extend as far as 1 Mpc from their host galaxy.}
\end{figure}

\end{document}